\documentclass[12pt]{iopart}

\usepackage{graphicx}
\usepackage{epstopdf}
\usepackage[usenames]{color}
\begin{document}

\title{1D momentum-conserving systems: the conundrum of anomalous versus normal
heat transport}

\newcommand{\CE}{C_E(i, t; j=0, t=0)}
\newcommand{\CP}{C_P(i, t; j=0, t=0)}

\author{Yunyun Li$^1$, Sha Liu$^2$, Nianbei Li$^1$$^*$, Peter
H\"anggi$^1$$^,$$^2$$^,$$^3$$^,$$^4$, and
Baowen Li$^1$$^,$$^2$$^,$$^5$}

\address{$^1$ Center for Phononics and Thermal Energy Science, School of Physics
Science and
Engineering, Tongji University, 200092 Shanghai, China}
\address{$^2$ Department of Physics and Centre for Computational Science and
Engineering, National
University of Singapore, 117456, Singapore}
\address{$^3$ Institut f\"ur Physik, Universit\"at Augsburg, Universit\"atsstr.
1, 86159 Augsburg,
Germany}
\address{$^4$ Nanosystems Initiative Munich, Schellingstr, 4, D-80799 M\"unchen,
Germany}
\address{$^5$ Center for Computational Science and Engineering, Graphene
Research Centre, Department
of Physics, National University of Singapore, 117456 Singapore}
\ead{nbli@tongji.edu.cn}
\date{\today}

\newcommand{\WARN}[1]{\textcolor{green}{#1}}
\newcommand{\NOTES}[1]{\textcolor{red}{#1}}
\begin{abstract}

Transport and  the spread  of heat  in  Hamiltonian one dimensional (1D)  momentum conserving nonlinear systems is commonly thought to proceed anomalously.
Notable exceptions, however,  do exist of which the coupled rotator model is a prominent
case.  Therefore, the  quest arises to identify the origin of manifest anomalous energy and momentum transport in those low dimensional systems.
We develop the theory  for both,  the statistical densities for momentum- and energy-spread and particularly its momentum-/heat-diffusion behavior, as well as its corresponding momentum/heat transport features.
We demonstrate that the second temporal derivative of the mean squared deviation
of the momentum spread is proportional  to the equilibrium correlation
of the total momentum flux.
Subtracting the part which corresponds to a ballistic momentum spread  relates (via this integrated, subleading momentum flux correlation) to an effective viscosity, or equivalently, to the underlying momentum diffusivity.
We next put forward the intriguing hypothesis: normal  spread of this so adjusted excess momentum  density   causes  normal
energy spread and alike normal  heat transport (Fourier Law).
Its corollary being that  an anomalous,  superdiffusive broadening of this adjusted  excess momentum density in turn implies
an anomalous energy spread and correspondingly  anomalous, superdiffusive heat transport.
This  hypothesis is successfully  corroborated within extensive molecular dynamics simulations over large
extended time scales. Our numerical validation of the hypothesis  involves four distinct
archetype classes of nonlinear pair-interaction potentials:  (i)
a globally bounded pair interaction (the noted coupled rotator model), (ii) unbounded interactions acting at large
distances
(the coupled rotator model amended with
harmonic pair interactions), (iii) the case of a hard point gas with unbounded square well interactions and  (iv) a pair interaction potential being unbounded at short distances while displaying an asymptotic free part (Lennard-Jones model). We compare  our findings with recent predictions obtained from nonlinear fluctuating hydrodynamics theory.

\end{abstract}
\maketitle
\normalsize


%


\section{Introduction}
The investigation of heat conduction in low dimensional nonlinear lattices has
attracted ever increasing
attention in the statistical physics
community \cite{Lepri2003pr,Dhar2008ap,Liu2013epjb}. Although early relevant work \cite{Kaburaki1993pla} can be traced back to $1993$,
an increased activity has spurred since the discovery of anomalous heat conduction
occurring in one dimensional (1D) momentum-conserving Fermi-Pasta-Ulam
(FPU)-$\beta$ lattices \cite{Lepri1997prl} in 1997. In those  low dimensional study
cases the thermal
conductivity $\kappa$ of the FPU-$\beta$ lattice was found to diverge with the
lattice size $N$ as $\kappa\propto N^{\alpha}$,
with $0< \alpha <1$.
This finding consequently  yields a system-size dependent thermal
conductivity, thus  breaking Fourier's law of heat conduction.
Similar anomalous heat conduction behavior has also been identified for other
archetype 1D momentum-conserving stylized nonlinear systems, such as the
1D diatomic Toda lattices \cite{Hatano1999pre}, and, importantly, has been
predicted to occur in momentum-conserving physical
materials, such as in carbon nanotubes \cite{Zhang2005jcp},
silicon nanowires \cite{Yang2010nt} and in polymer chains \cite{Liu2012prb}.
Experimentally, the
breakdown of Fourier's law has presently
been confirmed for 1D carbon nanotubes and boron-nitride
nanotubes \cite{Chang2008prl} and in 2D suspended graphene \cite{xu2014nc}.

On the other hand, the low (1D, 2D) spatial dimension alone
is not the sole  feature that determines whether the validity of Fourier's law
holds up. For example, normal heat conduction obeying Fourier's law
has been established beyond doubt for 1D nonlinear Frenkel-Kontorova (FK)
\cite{Hu1998pre} lattices and
$\phi^4$ lattices \cite{Hu2000pre,Aoki2000pla}. For those nonlinear lattice systems the
total momentum is not conserved, being  due to the
presence of the on-site potentials. These numerical results for 1D lattices led
to a conjecture that the property of momentum-conservation in low dimensional
systems might be at the origin to give rise to anomalous heat conduction for 1D
and 2D nonlinear lattices, e.g. see
\cite{Lepri2003pr,Dhar2008ap,Prosen2000prl,Narayan2002prl}.
It then later came as a surprise that contradictory results  emerged  for other stylized
momentum-conserved nonlinear 1D lattices, exhibiting
saturated thermal conductivities  such as the rotator model
\cite{Giardina2000prl,Gendelman2000prl} and a momentum-conserving variation of the
ding-a-ling model \cite{Dadswell2010pre}.  Giardin{\`a} and Kurchan also
provided a family of models with or without
momentum-conservation which, however, all obey Fourier's law \cite{Giardina2005jsat}.
Therefore this situation gives rise to the dilemma of what physics is at the
root for the occurrence of the breakdown of the Fourier behavior in 1D nonlinear
lattices \cite{Flach2003chaos,Li2005chaos}. Most recently, relying on numerical
simulations, Savin and Kosevich \cite{Savin2014pre} showed that thermal
conduction obeys Fourier's law for 1D momentum-conserving lattices with a 1D
Lennard-Jones interaction, a Morse interaction, and as well a Coulomb-like
interaction.  Those numerical findings let them to conclude (we think erroneously, see in  Sect. $4.4$ below,  and, as well,
in Ref. \cite{Das2014pre}) that normal heat conduction emerges for momentum-conserving lattices whenever  the pair interaction
potentials are asymptotically  free at large interaction distances.

In this work, we focus on heat transport in 1D momentum-conserving nonlinear
lattices from another aspect, namely, the {\it diffusive spread} of energy and momentum.
It is acknowledged that there exists a profound
connection between heat conduction and heat
diffusion within the region where Fourier's law is valid. For example, take the normal heat conduction
in 1D cases: Fourier's law states that  $j=-\kappa \partial_x{T}$,
where $j$ denotes the local heat flux and $\partial_x{T}$ is the nonequilibrium
temperature gradient. If we combine this with local energy conservation; i.e.,
$\partial_t{E}+\partial_x{j}=0$ and, additionally, use the
relation between the local energy density $E$ and the temperature $T$, i.e., $E=c_{\mbox{v}} T$
(with $c_{\mbox{v}}$ being the
volumetric specific heat), then the familiar heat diffusion equation
$\partial_t{T}=D\partial^2_{x}{T}$ can be derived. The normal heat diffusivity equals $D=\kappa/c_{\mbox{v}}$.

Microscopically, normal heat diffusion can be characterized by the mean square displacement
of the corresponding Helfand moment \cite{Helfand1960pr}, which then connects to normal heat conductivity
via the Green-Kubo formula. The efforts trying to
bridge heat
conduction and diffusion beyond the normal case have only been put forward in the recent decade
\cite{Li2005chaos,Das2014pre,Denisov2003prl,Zaburdaev2011prl,Zaburdaev2012prl,Dhar2013pre,Beijeren2012prl,Mendl2013prl,Spohn2014JSP,Mendl2014pre}.
Remarkably, it is only  recently
that a general and rigorous connection
between heat conduction and heat diffusion has been established from first
principles \cite{Liu2014prl}: It is shown that in the linear response regime,
the evolution of the second order time-derivative of the mean squared deviation (MSD) of
a general energy diffusion process is determined  by the equilibrium heat
flux autocorrelation function of the system -- the central quantity that enters the Green-Kubo formula for the thermal heat conductivity.
The key ingredient for obtaining this MSD of the energy spread relies  on
the energy-energy
correlation function $C_E(x,t;x',0)$ \cite{Zhao2006prl}, as
rigorously shown in recent work \cite{Liu2014prl}.
This thermal equilibrium excess energy-energy correlation indeed
is the fundamental quantity that determines the behavior of nonequilibrium heat
diffusion, as well as the nonequilibrium heat conduction  in a
regime not too far displaced from thermal equilibrium. Thus, using the energy-energy correlation
function, we can conveniently identify whether the heat diffusion in a nonlinear lattice occurs normal
or anomalous.

With this present study  we aim to shed more light on the conundrum that
underpins anomalous heat transport in 1D nonlinear lattices.
In doing so we study with molecular dynamics (MD) simulations four different
nonlinear 1D momentum-conserving nonlinear
lattices. The 1-st one is the 1D coupled rotator lattice which has a bounded
interaction potential; i.e., the
potential is bounded in configuration space and therefore the motion of the
particles are not confined. The 2-nd test case studies
an unbounded harmonic interaction potential in  combination with the coupled
rotator interaction potential. The 3-rd test case is  the hard point gas model with alternating masses
subject to  infinite square well pair interactions. This model is believed to show good
mixing properties and therefore fast convergence features.  As yet a 4-th 1D nonlinear system we
complement the rotator model with a Lennard-Jones 1D-interaction potential,
being unbounded at short interaction distances while being free at large
interaction distances. This latter model thus allows for bond dissociation at
large interaction distances. For all these test beds the correlation functions for the local excess
energy deviations as well as the local excess momentum are calculated via
extensive equilibrium numerical MD-simulations.

Our studies corroborate the result that  normal heat diffusion is found for the
coupled rotator lattice. We also demonstrate
that in addition to normal heat diffusion the overall dynamics is accompanied  by
a normal  momentum diffusion.
We then elucidate that  these two features imply that the system dynamics is
ruled  by the emergence of
a finite  momentum diffusivity.  This observation therefore insinuates that the
1D rotator model physically mimics a fluid behavior.  In clear contrast, we
find  that  anomalous heat diffusion occurs for momentum-conserving nonlinear 1D
lattices which contain an unbounded interaction potential, as it is the case
also with nonlinear FPU-lattices, the hard point gas and also the Lennard-Jones case. The anomalous
heat diffusion and corresponding anomalous heat conductivity behavior is shown
to be accompanied in all those test cases with the momentum excess density to undergo anomalous
superdiffusion. This latter feature  causes a
divergent effective viscosity, thus mimicking physically a solid-like behavior.

The present study  is organized as follows. In Section 2, we briefly review the
state of the art of the theory for excess energy diffusion and then develop the
theory describing the diffusion of excess momentum. In Section 3, we put forward our hypothesis for the occurrence of normal/anomalous heat transport. This
hypothesis is tested thoroughly in Section 4.  We start out by performing
numerical studies on an overall  bounded interaction potential, namely the coupled
rotator model. This is then followed  by studying a variant of this rotator
model by complementing it with  unbounded harmonic pair interactions. In addition we
discuss the cases with a hard point gas and  a Lennard-Jones pair interaction. These detailed
numerical MD studies  for these four nonlinear lattice systems support the fact that it is not the mere  presence or
absence of the symmetry  of momentum conservation but rather  the
presence  or absence of a fluid-like behavior, as characterized with normal spread of the momentum excess density,
which we speculate to be at the source  for the validity or the breakdown of
Fourier's law behavior. For the prior known cases with the dynamics subjected in addition to nonlinear on-site potentials
the momentum conservation is broken:  the emergence of Fourier's Law in this
latter situation is then ruled by nonlinear scattering processes which provide a
finite mean free path behavior for the heat transfer \cite{Liu2014prb}. Additional
conclusions and remaining open issues are presented with Section 5.

\section{Diffusion of heat and momentum}
Let us consider systems with a   momentum-conserving, homogeneous 1D nonlinear Hamiltonian lattice
dynamics with nearest neighbor interactions. Their Hamiltonian can be
cast in the general form
\begin{equation}\label{ham}
H=\sum_{i} \left[\frac{p_{i}^2}{2m} + V(q_{i+1}-q_{i})\right]\equiv\sum_{i} H_i
\;,
\end{equation}
where the set $p_i$ denote the momenta of particles of identical masses $m$. The set $q_i$ are the displacements from the
equilibrium position
for the $i$-th atom with $i=0, \pm1, \pm2,..., \pm (N-1)/2$, where an odd value of $N$ is assumed for the sake of convenience. The part $V(q_{i+1}-q_{i})$
is the interaction potential between neighboring sites $i$ and $i+1$. With $H_i$
we formally denote the
local energy at site $i$. Moreover, throughout  our numerical analysis  we shall
make use of periodic boundary conditions; i.e., we set $q_{N+i}=q_i$  and
$p_{N+i}=p_i$. The center of mass velocity of the system is chosen at rest; i.e. $v_{cm} =0$.
Note also that we use here strictly Hamiltonian lattice systems which contain no stochastic interaction parts of a spatial or a temporal nature.

\subsection{Heat diffusion}
We start out with the description of heat diffusion in a discrete 1D lattice
following Ref. \cite{Liu2014prl}.  In doing so, we introduce the energy-energy
correlation
function, reading:
\begin{equation}\label{e-corr}
C_E(i,t;j,0)\equiv\frac{\left<\Delta{H_i(t)}\Delta{H_j(0)}\right>}{k_B T^2 c_{\mbox{v}}},
\end{equation}
where $\Delta{H_i(t)}\equiv H_i(t)-\left<H_i(t)\right>$ and
$\left<\cdot\cdot\cdot\right>$ denotes
the ensemble average over canonical thermal equilibrium at a temperature $T$ and
$c_{\mbox{v}}$ is the
specific heat per particle.

Given this autocorrelation function of energy fluctuations, one can evaluate the
time evolution of the excess energy
distribution $\rho_E(i,t)$ starting out from an initial, near thermal
equilibrium state, characterized by the initial excess energy perturbation
$\xi(i)$. We consider the  case of a localized, small initial excess
energy perturbation at the central site, i.e.,  $\xi(i)= \epsilon \delta_{i,0}$.
We can then use linear response theory for the excess energy distribution
$\rho_E(i,t)$  to obtain \cite{Liu2014prl}:
\begin{equation}
\mkern-36mu \mkern-36mu \rho_E(i,t)= \sum_{j}C_E(i,t;j,0)\xi(j)/ \epsilon =\CE ,\,\,-\frac{N-1}{2}\leq i
\leq \frac{N-1}{2}.
\end{equation}
This excess energy distribution remains normalized at
all later times $t$, being due to the  conservation of energy.

The  commonly used quantity which quantifies the speed of {\it heat diffusion}
is the MSD
$\left<\Delta{x^2(t)}\right>_E$ of the
excess energy distribution. For a discrete 1D lattice with  $N$ sites one thus
obtains with
$\left<{x(t)}\right>_E=0$
\begin{equation}\label{heat-dif}
\mkern-36mu \mkern-36mu \left<\Delta{x^2(t)}\right>_E\equiv\sum_{i} i^2\rho_E(i,t)=\sum_{i} i^2
\CE{},\,\,-\frac{N-1}{2}\leq i \leq \frac{N-1}{2},
\end{equation}
 This MSD has
been  shown to obey the salient second order
differential equation \cite{Liu2014prl}; i.e.,
\begin{equation}\label{heat-theory}
\frac{d^2\left<\Delta{x^2(t)}\right>_E}{dt^2}=\frac{2}{k_B T^2 c_{\mbox{v}}}C_J(t)\;,
\end{equation}
where $C_J(t)$ denotes the equilibrium autocorrelation function of total heat flux defined
as
\begin{equation}
C_J(t)=\frac{1}{N}\left<\Delta J(t)\Delta J(0)\right>,\,\,J(t)=\sum_i{j_i}\;,
\end{equation}
wherein $j_i\equiv -\frac{p_i}{m}\partial{V(q_{i}-q_{i-1})}/\partial{q_i}$ is the local
heat flux. Note that this correlation $C_J(t)$ is  just what enters the Green-Kubo formula for thermal conductivity \cite{Lepri2003pr,Dhar2008ap,Visscher1974pra,Allen1993prb},  being  written as $\left<J(t)J(0)\right>/N$. This is so because here with $v_{cm}=0$ and $\Delta J(t) = J(t)$, as the equilibrium average obeys $\left<J(t)\right>=0$. Moreover, $J(t)$ contains no energy current stemming from transporting charge in an electromagnetic field or  an energy  current stemming from a particle concentration gradient.

$C_J(t)$ is the quantity that  enters the well-known Green-Kubo
expression for the thermal conductivity $\kappa$. For
normal heat flow it explicitly reads,
$\kappa=1/(k_BT^{2})\int_{0}^{\infty}C_J(t)dt$.

The relation in (\ref{heat-theory}) connects heat conduction with heat diffusion
in a rigorous way.
As a consequence,  the investigation of heat conduction can  equivalently be
obtained from studying
heat diffusion. The most important quantity is the energy fluctuation
autocorrelation function $\CE$ in Eq.
(\ref{e-corr});  it encodes all the necessary information about heat diffusion
and heat conduction.
As one can defer from Eq. (\ref{heat-dif}) and Eq. (\ref{heat-theory}), the
energy-energy correlation
function $\CE{}$ determines the dynamical behavior of the MSD of heat diffusion
as well as
the autocorrelation function of total heat flux $C_J(t)$.

As an example take the FPU-$\beta$ model which displays
anomalous heat diffusion: there, the energy autocorrelation
$\CE$ follows a Levy walk distribution, being quite distinct from a normal
Gaussian distribution in the long time
limit \cite{Zaburdaev2011prl,Zaburdaev2012prl,Zhao2006prl}. This statistics then  gives rise to
a superdiffusive behavior for  the  energy spread, reading
\begin{equation}
\left<\Delta{x^2(t)}\right>_E\sim t^{\beta},\,\,1<\beta<2 \;.
\end{equation}
The corresponding, formally diverging anomalous thermal conductivity
can be extracted to read \cite{Liu2014prl}
\begin{equation}
\kappa\sim
\frac{1}{k_B
T^2}\int_{0}^{N/c}C_J(t)dt=\left.\frac{c_{\mbox{v}}}{2}\frac{d\left<\Delta{x^2(t)}
\right>_E}{dt}\right|_{
t\sim
N/c}\propto N^{\beta-1}\;.
\end{equation}
Here,  $t_s\sim N/c$ with $N$ chosen sufficiently large presents  the
characteristic time-scale of heat diffusion.  The quantity $c$ refers to the speed of sound
for inherent renormalized phonons \cite{Li2010prl}.

\subsection{Momentum diffusion}
The scheme for the excess energy heat diffusion can likewise be generalized for
the problem of corresponding diffusion of excess momentum.
For a nonlinear lattices with a Hamiltonian in  Eq. (\ref{ham}), the
translational
invariance of the Hamiltonian necessarily indicates that the total momentum
$\sum_i p_i$ is
conserved; i.e., we have
\begin{equation}
\frac{d\sum_i
p_i}{dt}=-\sum_i\left(\frac{\partial{V(q_i-q_{i-1})}}{\partial{q_i}}-\frac{
\partial{V(q_{i+1}-q_{i})
}}{\partial{q_{i+1}}}\right)=0 \;,
\end{equation}
by observing that
$\partial{V(q_{i+1}-q_{i})}/\partial{q_i}=-\partial{V(q_{i+1}-q_{i})}/\partial{
q_{i+1}}$.

Using an analogous reasoning as put forward with the preceding subsection for
heat diffusion we can  define the autocorrelation function for the excess
momentum fluctuation \cite{Zhao2006prl}, reading explicitly:
\begin{equation}\label{p-corr}
C_P(i,t;j,0)=\frac{\left<\Delta{p_i(t)}\Delta{p_j(0)}\right>}{mk_B T},
\end{equation}
where $\Delta{p_i(t)}\equiv p_i(t) -
\left<p_i(t)\right>= p_i(t)$, observing that $\left<p_i(t)\right>=0$ in thermal
equilibrium.
Following the reasoning of the previous subsection we next demonstrate that
this momentum-momentum autocorrelation function describes, within linear response
theory, the
diffusion of momentum along the lattice.

To elucidate this issue we consider alike a lattice in thermal
equilibrium at temperature
$T$. We apply a small  kick of short duration to the $j$-th particle.
The kick occurs
with a constant impulse $\mathcal{I}$, yielding a force kick  at site $j$ as
\begin{equation}
	f_j(t)= \mathcal{I}\delta(t).
\end{equation}
Upon integrating the equation of motion from the moment immediately before the
kick (denoted as
$t=0^-$) to the moment immediately after the kick (denoted as $t=0^+$), we find
that the sole effect
of this kick is to change the momentum of the $j$th particle by an amount
$\mathcal{I}$. The
momenta of all other particles, as well as the position of all particles remain
unchanged. Formally,
this is recast as
\begin{equation}
\label{eq:dp}
  p_i(t=0^+)-p_i(t=0^-)= \mathcal{I} \delta_{i,j};
 \end{equation}
 \begin{equation}
  q_i(t=0^+)-q_i(t=0^-)= 0.
\end{equation}

The full time evolution of the momenta and positions is not analytically
accessible for
non-integrable nonlinear lattice systems. However, given that $\mathcal{I}$ is
small,  the validity regime
of linear response is obeyed.  The explicit response can be obtained by
referring to canonical linear response
theory for an isolated system \cite{Hanggi}.  Specifically, we assume that the
system has been prepared in the infinite
past, $t=-\infty$, with the canonical distribution
\begin{equation}
	\rho(t=-\infty)=\rho_{eq}= \frac{1}{Z} \exp[-\beta_T H]; \quad Z=\int
d\Gamma \exp[-\beta_T
H]\;,
\end{equation}
where $\beta_T=1/k_BT$ and $d\Gamma= dq_1\cdots dp_1\cdots$. With a time
dependent force $f_j(t)$
applied to the $j$th
particle, the total Hamiltonian reads $H_{tot}= H- f_j(t) q_j$. With the system
dynamics being closed, the evolution of the phase space
distribution is governed by the Liouville equation
\begin{equation}
	\frac{\partial \rho(t)}{\partial t}=\{ H_{tot}, \rho(t)\} \equiv
L_{tot}\rho(t)\;,
\end{equation}
where $\{\cdots,\cdots\}$ denotes the Poisson bracket. The linear response
solution
can be readily obtained up to the first order of $f_j$, yielding

\begin{equation}
\rho(t)= \rho_{eq}+ \Delta \rho(t)= \rho_{eq}+ \frac{1}{m k_B T} \int_0^\infty
ds e^{L s} p_j
\rho_{eq} f_j(t-s),
\end{equation}
The operator $L$ is the  Liouville operator  for the original, unperturbed
system, i.e. $LA=\{
H, A\}$
for any quantity $A$.
Therefore, in presence of the kick-force  the thermally averaged particle
momenta read for $t>0$
\begin{equation}
\left<p_i(t)\right>_{response}= \int p_i \Delta \rho(t) d\Gamma
=\frac{\mathcal{I}\left<\Delta p_i(t)\Delta p_j(0)\right>}{mk_B T}=
\mathcal{I}C_P(i,t;j,0).
\end{equation}
For $t=0^+$, it reduces to
$\left<p_i(0^+)\right>_{response}= \mathcal{I}\delta_{i,j}$ due to equipartition $\left<p_i(0)
p_j(0)\right>=m k_B T\delta_{i,j}$, which is
consistent with Eq.
(\ref{eq:dp}).

The conservation of total momentum  implies that,
$\sum_i\left<p_i(t)\right>_{response}$,
is  conserved as well. Evaluating this sum
at time $t=0$ yields $\sum_i C_P(i,t; j,0)= 1$ for all later times $t$. The
excess momentum  density function $\rho_P(i,t)$ therefore assumes the form
\begin{equation}
\rho_P(i,t)= \frac{\left<p_i(t)\right>_{response}}{\sum_i
\left<p_i(t)\right>_{response}}=C_P(i,t;j,0),
\end{equation}
which remains normalized in the course of time $t>0$. The density $\rho_P(i,t)$ is, however,  not
necessarily semi-positive everywhere; i.e. it again does not present a manifest probability density for all later
times $t$.

With time evolving, we notice that
the excess momentum autocorrelation Eq. (\ref{p-corr})   describes the
spread of
the momentum distribution after the initial kick has occurred. As can be
observed below, for
increasing times $t$ the quantity $C_P(j,t; j,0)$
decreases (at least for some finite time). This implies the decrease of the
momentum of the
$j$'th particle. The lost momentum is transferred to its neighbors. This feature
physically mimics a viscous behavior.

Let us next assume that the kick is applied to the center particle; i.e. we
explicitly set
$j=0$. Similarly to Eq.
(\ref{heat-dif}), we define the MSD of the excess {\it momentum }
$\left<\Delta{x^2(t)}\right>_P$ for a discrete lattice as
\begin{equation}
\mkern-36mu \mkern-36mu \left<\Delta{x^2(t)}\right>_P=\sum_{i} i^2 \rho_P(i,t)= \sum_i i^2
\CP,\,\,-\frac{N-1}{2}\leq i \leq
\frac{N-1}{2}\;.
\end{equation}

Because of the conservation of total momentum, in analogy to the energy
continuity relation, we
may define a ``momentum flux'' $j^P_i$  via the local momentum continuity
relation. To see this, we
write down the Newtonian equation of motion for the $i$'th particle, reading
\begin{equation}
\frac{d p_i}{d t}=
-\frac{\partial{V(q_i-q_{i-1})}}{\partial{q_i}}-\frac{\partial{V(q_{i+1}-q_{i})}
}{\partial{q_{i}}}.
\end{equation}
By defining the momentum flux as
$j^{P}_i=-\partial{V(q_i-q_{i-1})}/\partial{q_i}=
\partial{V(q_i-q_{i-1})}/\partial{q_{i-1}}$, we obtain a discrete form of the
momentum continuity relation, reading
\begin{equation}
	\frac{d p_i}{d t}-j_i^P+j_{i+1}^P=0\;.
\end{equation}
Note that the momentum flux $j_i^P$ is actually the force exerted on particle
$i$ from particle
$(i-1)$. Its ensemble average $\left<j^{P}_i\right>$ yields the average
internal pressure.

Following the strategy  used  for heat diffusion, one can
derive a corresponding
relation for the second time derivative $\left<\Delta{x^2(t)}\right>_P$. It reads:
\begin{equation}
\frac{d^{2}\left<\Delta{x^2(t)}\right>_P}{dt^2}=\frac{2}{mk_BT}C_{J^{P}}(t) \;.
\label{eq:relation}
\end{equation}
Here,  the centered autocorrelation function of the momentum flux is given by
\begin{equation}
C_{J^{P}}(t)=\frac{1}{N}\left<\Delta J^{P}(t)\Delta J^{P}(0)\right>,\,\,J^{P}=\sum_{i}j^{P}_i\;.
\end{equation}

 It should be observed that  here the momentum flux $\Delta J^P(t)$, unlike for energy flux,  cannot be replaced with $J^P(t)$ itself. This is so  because the equilibrium average is  typically non-vanishing with $\left<J^P(t)\right> = N \Lambda$, where ${\Lambda}$ denotes a possibly non-vanishing internal equilibrium pressure in cases where the interaction potential is not symmetric.

The presence of a finite, isothermal  sound speed $c$ may imply that the momentum spread contains  a ballistic component. Spreading then occurs into the positive and negative directions with velocity $c$, with the two centers of equal weight $1/2$ moving at velocities $\pm {c}$ \cite{Helfand1960pr}. We hence must subtract this trivial ballistic part $\frac{1}{2}c^2 t^2$ for the weighted $(\frac{1}{2})$ one-sided spread in configuration space.  The effective bulk viscosity $\eta$  is thus given as an   integration over this {\it subleading} excess momentum correlation $C_{J^{P}}(t)$ over time  in terms of a Green-Kubo formula \cite{Helfand1960pr,Resibois}, reading

\begin{equation}\label{eta}
\eta\equiv \lim_{t\rightarrow \infty}\Big(\frac{1}{k_B T} \int_0^{t} C_{J^{P}}(t)dt - \frac{1}{2}mc^2 t\Big).
\end{equation}
In case that the momentum diffusion occurs normal  one can invoke the  concept
of a  finite momentum
diffusivity by defining, upon use of eqs. (\ref{eq:relation}, \ref{eta}):

\begin{equation}
\label{eq:D_P}
 2 D_P\equiv \lim_{t\rightarrow \infty}\Big(\frac{d\left<\Delta{x^2(t)}\right>_P}{dt} - c^2 t\Big)
\; .
\end{equation}
Therefore, for the discrete lattices discussed here, this so introduced
viscosity $\eta$ precisely
equals the momentum diffusivity times the atom mass, namely
\begin{equation}
	\eta= m D_P \;.
\label{eq:eta-D_P}
\end{equation}
Given a situation where the excess momentum density spreads not normally  the
limit in
Eq. (\ref{eq:D_P}) no longer exits. The integration in Eq. (\ref{eta}) formally
diverges, thus leading to an infinite viscosity.

In the context of this work we find that such an infinite viscosity
indicates a manifest solid-like behavior.  In distinct contrast, however, a
result with a   finite effective viscosity indicates an effective  fluid-like behavior.

\section{The hypothesis}
The general  folklore in the field of anomalous heat conduction
\cite{Prosen2000prl,Narayan2002prl} is that in  momentum-conserving 1D nonlinear
lattices one encounters an anomalous heat conductance behavior. The case with
the rotator model, however, presents  an eminent exception.
So what is the physical mechanism which can explain such exceptions? -- An
observation is that
in all those presently known cases exhibiting anomalous 1D heat conductance the
interaction potential has been of
unbounded nature at large interaction distances. The known exceptions,
predominantly the well studied case with the rotator model,  do not possess such
unbounded pair interactions at long distances. Obviously the form of the
overall interaction does matter for the violation of Fourier's law.
One may speculate that  the emergence of the anomalous behavior is rooted in the form of an excess
momentum density dynamics that behaves  solid-like in the sense that the momentum
diffusion does not support a finite effective  viscosity in the spirit defined above. In contrast, a
Fourier-like behavior may become possible if the inherent momentum dynamics is
more fluid-like,
consequently possessing a finite effective momentum diffusivity.
An appealing conjecture therefore is that it is the physics of  momentum
diffusion which rules whether  heat transport occurs normal
or anomalous. In short, we next test with different models the following hypothesis:\\

(i) {\it Heat transport in nonlinear 1D momentum-conserving Hamiltonian lattice systems occurs normal
whenever the spread of the profile of the  excess momentum density, upon subtracting a possibly present leading ballistic part,  is normal.} \\

(ii) {\it The corollary being that heat transport occurs anomalous whenever this so adjusted, subleading momentum excess density spreads superdiffusive}.\\

If this hypothesis holds true it is expected to hold vice versa, i.e., with heat/momentum substituted by momentum/energy.

\section{Testing the hypothesis}
We next test this so stated hypothesis numerically with four classes of nonlinear Hamiltonian lattice dynamics.
The numerical procedure used and the details of scaling of parameters and dimensionless units are deferred to the Appendix.

\subsection{Coupled rotator dynamics}
As a first test bed for the above hypothesis  we scrutinize the   normal
heat transport behavior in a nonlinear, momentum-conserving 1D
occurring with the coupled rotator lattice. Throughout the remaining  we shall use Hamiltonian lattice models  with corresponding dimensionless units \cite{Lepri2003pr,Dhar2008ap}.  The  Hamiltonian   for the coupled rotator lattice dynamics reads
\begin{equation}\label{Ham-con}
  H =  \sum_{i} \Big( \frac{ p_{i}^2}{2} + [1- \cos(q_{i+1} - q_{i})]\Big) \;.
\end{equation}
Notably, here the nonlinear, momentum-conserving interaction potential is
bounded for all arguments
via the cosine function. The local energy density is
$H_{i}=p_{i}^2/2 + [1- \cos(q_{i+1} - q_{i})]$. Without loss of generality, we
consider the
initial distribution of the excess energy or momentum to be a Kronecker-delta
function  in the lattice
center. The autocorrelation functions $\CE$ and $\CP$
for energy and
momentum are defined according to Eqs. (\ref{e-corr}) and (\ref{p-corr}). Thus,
the
temporo-spatial behavior of $\CE$ and $\CP$ describe the dynamics of energy and
momentum diffusion starting out from the central position. With the interaction potential being symmetric there is vanishing internal pressure.

\begin{figure}[t!]
  \centering
\includegraphics[width=0.40\textwidth]{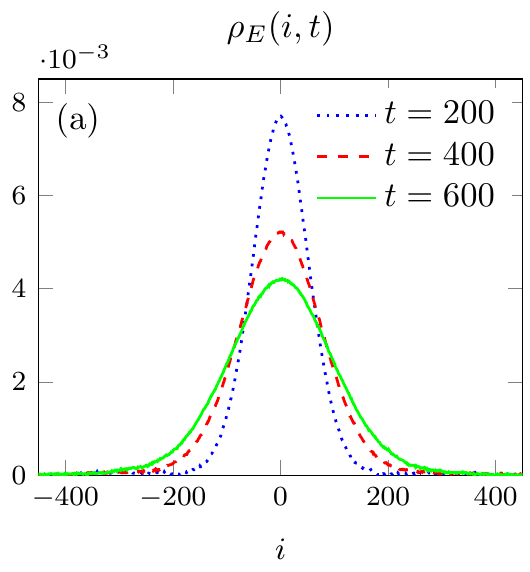}
\includegraphics[width=0.41\textwidth]{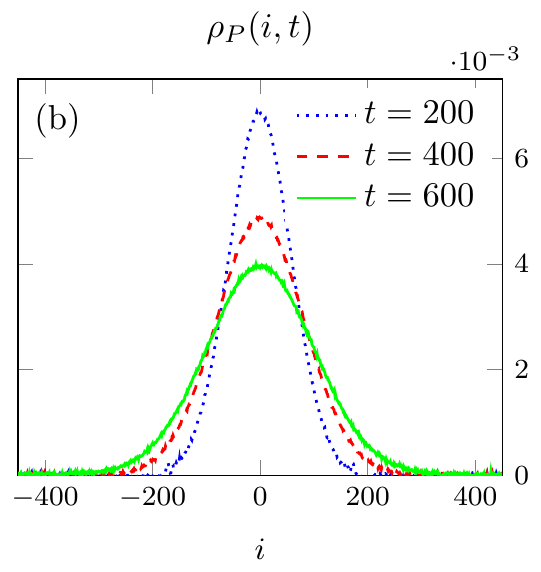}
\includegraphics[width=0.41\textwidth]{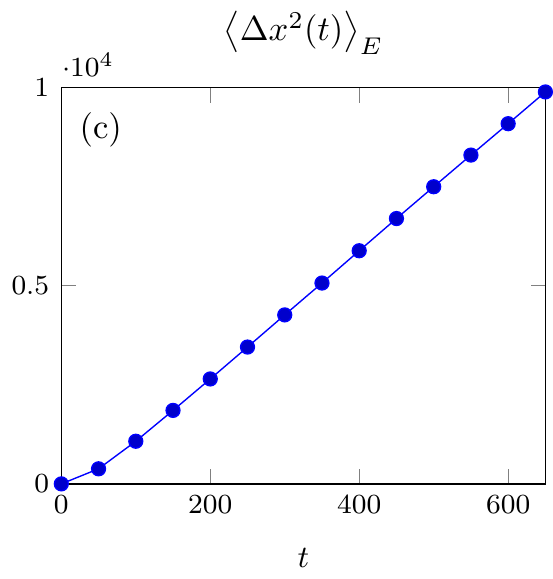}
\includegraphics[width=0.42\textwidth]{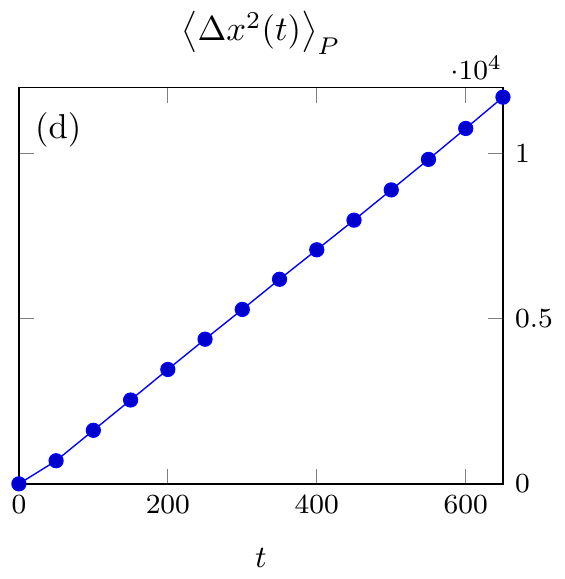}
\centering
\caption{(color online) Heat and momentum transport in the coupled rotator model: Upper panels (a) and (b):
Spatial distribution of the energy autocorrelation $\rho_E(i,t)=\CE$ and the momentum autocorrelation
$\rho_P(i,t)=\CP$, respectively. The correlation times
are $t=200$
(dotted blue),
$400$ (dashed red), and $600$ (solid green). Lower panels (c) and (d): The  mean squared deviation (MSD)
of the energy
 $\left<\Delta{x^2(t)}\right>_E$
and the momentum  $\left<\Delta{x^2(t)}\right>_P$, respectively. A
perfect linear time dependence
of the MSD can be clearly detected for both, the energy and the momentum.
 The lattice size is chosen
$N=1501$ and the temperature is $T\approx 0.413$.
}
  \label{corr-con}
\end{figure}

In Fig. \ref{corr-con} (a), we depict the  correlation functions $\CE$ for the
energy diffusion
versus evolving relative time span $t$.  For sufficiently large times  $t$ we
observe that the
energy autocorrelation
function $\CE$ evolves very closely into a Gaussian distribution function (but
still spatially bounded with the causal cone, as determined by a finite speed of
sound); i.e., its profile is perfectly well given by
\begin{equation}\label{gaussian}
C_{E}(i,t)\sim \frac{1}{\sqrt{4\pi D_{E} t}}e^{-\frac{i^2}{4D_{E} t}}
\end{equation}
with $D_E$ denoting the diffusion constant for heat diffusion. As a result, the
MSD of heat
diffusion $\left<\Delta{x^2(t)}\right>_E$ then  depicts at for sufficiently long
time $t$ a linear dependence in time $t$, being the hall mark for normal
diffusion.

In summary, normal diffusion for heat is accurately corroborated numerically
 with the findings depicted with  Fig.
\ref{corr-con} (c).
\begin{equation}
\left<\Delta{x^2(t)}\right>_{E}\sim\sum_{i}i^2C_{E}(i,t)=\sum_{i}i^2
\frac{1}{\sqrt{4\pi D_{E}
t}}e^{-\frac{i^2}{4D_{E} t}}= 2D_{E} t\;.
\end{equation}
Accordingly,  heat diffusion theory in \cite{Liu2014prl} for  normal diffusion
of heat
$\left<\Delta{x^2(t)}\right>_{E}$ implies  that the  heat conduction
behavior is normal as well, with the heat conductivity
given by $\kappa= c_{\mbox{v}} D_E$.

This Gaussian behavior for $\CE$ with its corresponding linear time-dependence
of the MSD for
heat diffusion
$\left<\Delta{x^2(t)}\right>_E\propto t$ has been observed previously in
nonlinear 1D lattices
which explicitly do  break momentum conservation by including an on-site
potential.  For example, this is so
for the case of  1D lattices with a $\phi^4$ on-site potential
\cite{Zhao2006prl}. In the latter
case it is agreed among all practitioners that normal heat conduction occurs
beyond any doubt
\cite{Hu2000pre,Aoki2000pla}.  The situation with  momentum-conserving
1D-coupled rotator lattices, however, is far from being  settled  in the
literature
\cite{Flach2003chaos,Li2005chaos}.  Here the possibility for  a diverging
thermal conductivity in the
thermodynamic limit is still considered as an option by some practitioners. The
present state of the art is nonconclusive although
prior extensive  numerical simulations, using either the Green-Kubo method or
the Non-Equilibrium Molecular Dynamics (NEMD) method, both seem to indicate that
the thermal
conductivity is size-independent \cite{Giardina2000prl,Gendelman2000prl}. The
source of the  ongoing dispute is that
the numerical results stemming either from the Green-Kubo method and/or the NEMD
method, all performed  for finite lattice sizes, may possibly not be
consistent with manifest asymptotic results in the thermodynamical limit.

In contrast, as we emphasized with the previous section, the energy
autocorrelation function $\CE$ constitutes a
fundamental, detailed measure yielding information well beyond  the MSD of
energy spread
$\left<\Delta{x^2(t)}\right>_E$ \cite{Liu2014prl, Li2010prl}. This is so because
of its  equivalence with the
Green-Kubo formula, which derives from the salient relation detailed with   Eq.
(\ref{heat-theory}).
Put differently, the
temporal-spatial distribution of $\CE$ yields improved, more detailed insight as
compared to a method that
merely evaluates via MD directly the Green-Kubo integral expression.

Next we study  the diffusion of the excess momentum via the
momentum
autocorrelation function $\CP$. Our findings are depicted with
Fig. \ref{corr-con} (b). One finds that not only does the energy diffusion obey
a Gaussian behavior,
but also the momentum diffusion occurs Gaussian within our explored
large regimes of
correlation time spans $t$.

 This   behavior of $\CP$ in this coupled rotator
lattice possessing a bounded interaction potential is
therefore very distinct from  the behavior of the
$\CP$ occurring in the momentum-conserving in FPU-$\beta$ lattice
\cite{Zhao2006prl}. Our
MSD of the excess momentum  $\left<\Delta{x^2(t)}\right>_P$ nicely follows a
perfect linear time
dependence, as can be  deduced from Fig. \ref{corr-con} (d).

According to Eq. (\ref{eta}), the viscosity $\eta$ for this coupled rotator 1D
lattice is therefore
finite. Put differently,  it exhibits a fluidic-like characteristics referred to
in the previous section. In distinct
contrast, the effective viscosity $\eta$
for the FPU-$\beta$ lattice is diverging towards infinity in the thermodynamic
limit; thus
displaying the  solid-like  characteristics, as  discussed in  section 2, cf.
see Eq. (\ref{eta}). In contrast to the case of the FPU-$\beta$ lattice with  three local conservation laws, here the angle
$(q_{i+1} - q_{i})$ is not conserved.  Thus, only two local conservation laws for momentum and energy are present, but none for the stretch  (or mass).
Nonlinear  fluctuating hydrodynamics theory then predicts a central, diffusive  spreading for momentum \cite{SpohnPC} without opposite moving side-peaks; -- this being in full agreement with our findings. The investigation of the {\it momentum} diffusion behavior in this coupled rotator lattice  (for a preliminary account see in the arXiv \cite{yunyun}) has inspired renewed  attention from other groups as well \cite{Spohn2014arxiv,Dhar2014arxiv}.

\subsection{Coupled rotator dynamics amended with harmonic interactions}

\begin{figure}
  \centering
\includegraphics[width=0.40\textwidth]{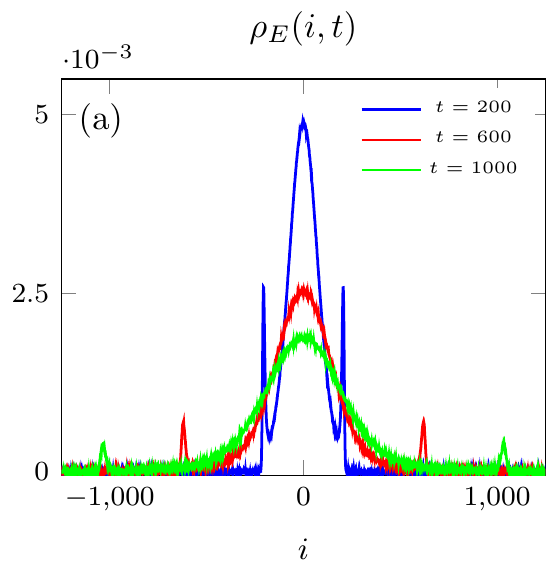}
\includegraphics[width=0.405\textwidth]{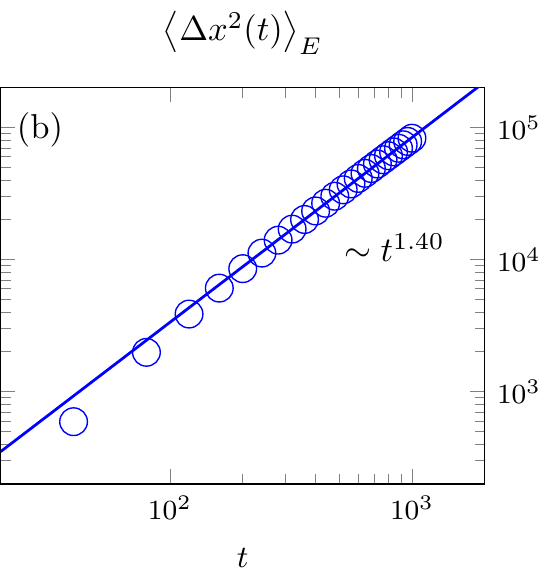}
  \caption{(color online) Heat transport in the amended rotator model with additional
harmonic pair interactions: (a): The normalized correlation
functions of excess energy density $\rho_E(i,t)=\CE$  for the rotator with unbounded interaction potentials. The
correlation time are
$t=200$
(blue),
$600$ (red), and $1000$ (green). (b): The MSD
of the energy spread
 $\left<\Delta{x^2(t)}\right>_E$. The time dependence ceases to be
linear for the energy diffusion. The solid blue power law lines serve as
a guide to the eye for the power law like behavior of the data in the large time
regime.  The parameters used
in the numerical simulations are $N=2501$ and $K = 0.5$. The calculated
equilibrium temperature is at $T\approx 0.800$.}
  \label{corr-noncon-energy}
\end{figure}

In testing  our hypothesis further we next amend the rotator coupling by adding
an additional  unbounded, but symmetric  harmonic interaction
potential. This transforms the original coupled rotator 1D lattice with bounded
interaction  into a
momentum-conserving 1D lattice with a vanishing internal pressure, but now with an  unbounded
pair interaction, being provided by the harmonic contribution. The
Hamiltonian for this so amended coupled rotator model reads:
\begin{eqnarray}
\label{Hamiltonian}\label{ham-ar}
  H=  \sum_{i} \left(\frac{p_{i}^2}{2} + [1- \cos(q_{i+1} - q_{i})]+
\frac{K}{2}(q_{i+1} -
q_{i})^2\right) \,,
\end{eqnarray}
where $K$ denotes the strength of the harmonic interaction. The total momentum
is  still
conserved.

Using the same numerical procedure  we  numerically study the  heat and momentum
diffusion for this set up. In
Fig. \ref{corr-noncon-energy} (a), the energy autocorrelation function $\CE$ at
different correlation times
is shown. The finite broadened side peaks exhibited by $\CE$ imply that heat
conduction no
longer proceeds normal; instead an anomalous,  faster--than--linear
superdiffusive time
dependence of the MSD of the energy spread
 $\left<\Delta{x^2(t)}\right>_E$ is depicted with Fig.
\ref{corr-noncon-energy} (b). This numerically
confirms that  heat conduction in
this unbounded 1D lattice is rendered anomalous. Our numerical fit exhibits
this superdiffusive
heat spreading, growing as $\left<\Delta{x^2(t)}\right>_E\propto t^{1.40}$. Notably, this superdiffusion exponent, $\beta=1.40$, for the amended rotator model is consistent with a previous result of $\beta=1.40$ for the FPU-$\beta$ lattice \cite{Zhao2006prl}.  Both, the amended rotator model and the FPU-$\beta$ lattice dynamics dwell a symmetric potential with a corresponding internal vanishing pressure.

We  emphasize that the energy autocorrelation function $\CE$ is directly connected with the transport coefficient of thermal conductivity \cite{Liu2014prl}.
In the recent developed {\it Nonlinear Fluctuation Hydrodynamic Theory}  (NFHT) \cite{Mendl2013prl,Spohn2014JSP}, three normal modes, including one central heat mode $f_0(x,t)$ and two opposite moving sound modes $f_{\pm 1}(x,t)$ are obtained upon expanding the  three Euler equations  up to second order only \cite{Spohn2014JSP}. Whether such a minimal modification is sufficient to model the transport features  is still under debate. In particular, it remains to be shown whether this approximate  procedure yields in fact a sufficiently good  approximation of the true dynamical transport behavior. In this spirit we hope that our present work sheds more light onto this still open question.

According to Spohn \cite{Spohn2014JSP}, the energy autocorrelation function $C_E(x,t)$ can be decomposed into the three normal modes as $C_E(x,t)=a f_{-1}(x,t)+b f_0(x,t)+a f_{+1}(x,t)$. The prefactors $a$ and $b$ are model dependent and usually depend on temperature. For example, it is obtained that $a=0$ and $b=0.83$ for the FPU-$\beta$ lattice at $T=1$ \cite{Das2014pre}. In this case, the energy autocorrelation function $C_E(x,t)$ and the heat mode $f_0(x,t)$ are equivalent, except for a different value for the prefactor. Therefore, the MSD obtained from the energy autocorrelation function $C_E(x,t)$ and of the central heat mode $f_0(x,t)$  should follow the same time dependence. However,  NFHT predicts an exponent of $\beta=1.50$ for the heat mode in lattices with symmetrical potential at zero pressure \cite{Spohn2014JSP}. This prediction for $\beta=1.50$, although quite close, distinctly differs nevertheless from   our finding here that $\beta=1.40$. This value $\beta=1.40$  agrees, as mentioned above, also with the  prior  results for the FPU-$\beta$ lattice dynamics \cite{Zhao2006prl,Wang2011epl}.

This discrepancy between the numerical results and the NFHT may originate from an apparent inconsistent assumption employed in Ref. \cite{Spohn2014JSP}: Namely, in Ref. \cite{Spohn2014JSP}, it is assumed that all the three peaks of the normal modes have a width much less than $c t$, where $c$ denotes the sound velocity. Using  this assumption, one employs the decoupling that the product $f_0(x,t)f_{\pm1}(x,t)\simeq 0$ for large $t$. Imposing such zero overlap one proceeds in  deriving that the diffusion of the sound modes occurs normal while the diffusion of heat mode is superdiffusive with an exponent of $1.50$. Note however that here the width of the heat mode ($\propto t^{1.50}$) exceeds $c t$ in the asymptotic large time limit, apparently thus contradicting the assumption made.

\begin{figure}
\centering
\includegraphics[width=0.40\textwidth]{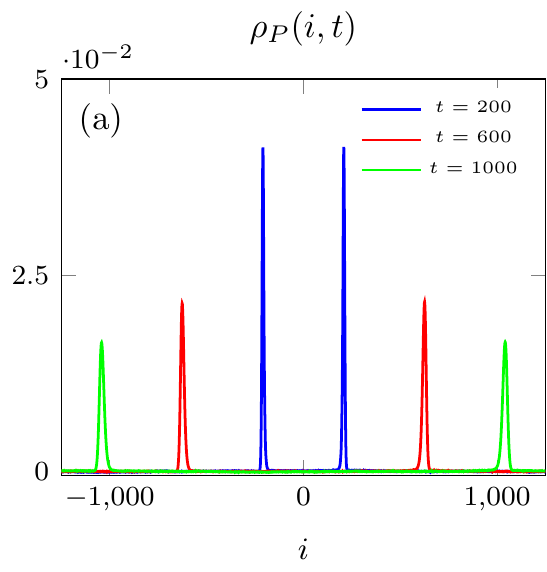}
\includegraphics[width=0.42\textwidth]{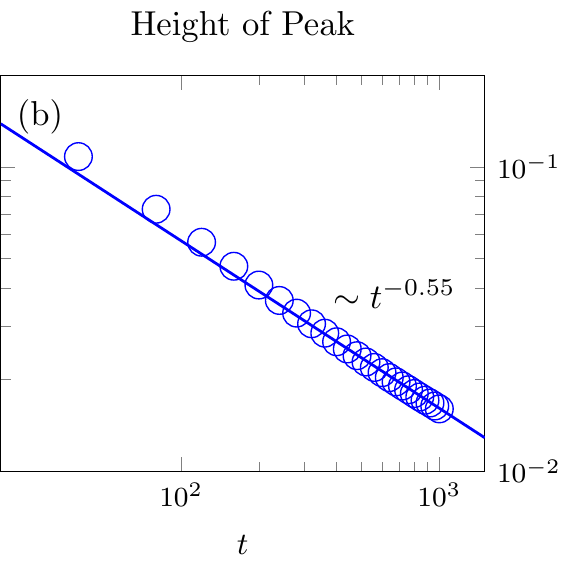}
\includegraphics[width=0.40\textwidth]{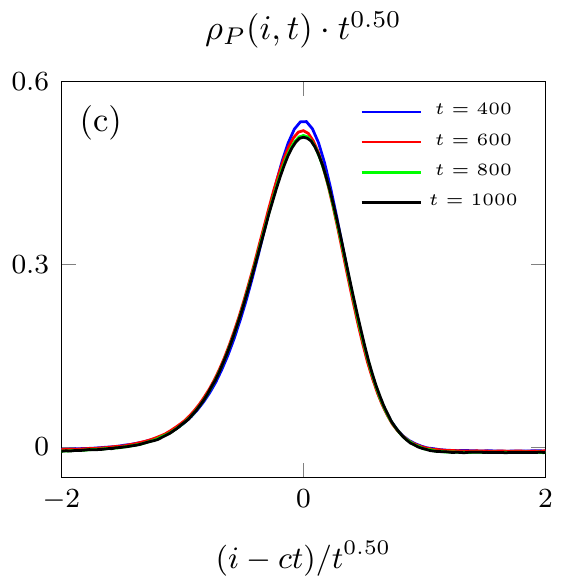}
\includegraphics[width=0.406\textwidth]{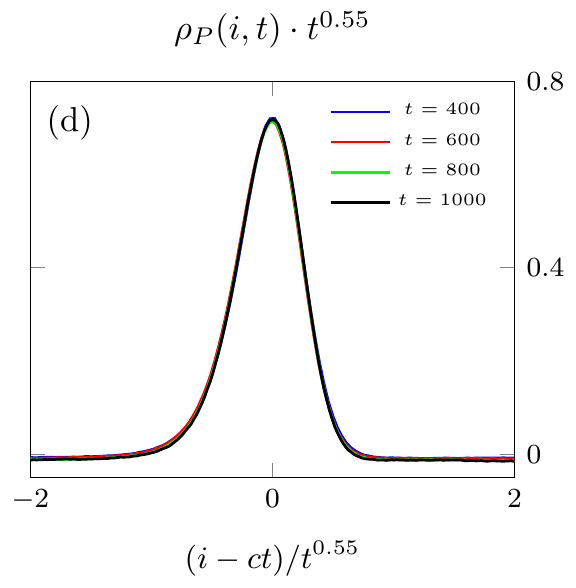}
\caption{(color online) Excess momentum  spread in the amended coupled rotator model with additional
harmonic pair interactions present: (a): The normalized correlation
functions of excess momentum density $\rho_P(i,t)=\CP$  for the rotator with unbounded interaction potentials. The
correlation time are
$t=200$
(blue),
$600$ (red), and $1000$ (green). Each has two symmetric side peaks moving outside with a constant sound velocity $c$. (b): The decay of the height of the side peak of $\rho_P(i,t)$. The solid blue power law lines with the dependence of $\sim t^{-0.55}$ is the best fit for the data from $t=400$ to $t=1000$. (c) The rescaled plot of the side peaks of $\rho_P(i,t)$ with the exponent of $0.50$ in the moving frame of sound velocity $c$ at $t=400,600,800$ and $1000$. (d)  The rescaled plot of the side peaks of $\rho_P(i,t)$ with the exponent of $0.55$ in the moving frame of sound velocity $c$ at $t=400,600,800$ and $1000$. The parameters used in the numerical simulations are $N=2501$ and $K = 0.5$. The calculated
equilibrium temperature is at $T \approx 0.800$.}
\label{corr-noncon-momentum}
\end{figure}

The question then arises whether this anomalous heat transport behavior is also reflected by the behavior for momentum diffusion. The numerically evaluated momentum autocorrelation function $\CP$ {\it vs.} the lattice site is depicted in Fig. \ref{corr-noncon-momentum} (a) for different correlation times. The solely present two side peaks move outwards with a constant sound velocity $c$, giving rise to a ballistic diffusion behavior for the momentum autocorrelation function $\CP$ with the leading term proportional to $c^2t^2$. The true diffusion behavior of momentum is reflected by the subleading term or the self-diffusion of the side peaks themselves \cite{SpohnPC}. The best way to illustrate this momentum behavior of self-diffusion is to present the decay of the height of the side peaks as a function of time. For a normal diffusion behavior this decay of the height of the peaks must follow an inverse square root law,  being proportional to $t^{-0.5}$. A decay faster than $t^{-0.50}$ does manifest itself as a non-diffusive, superdiffusive behavior. Indeed this feature is corroborated numerically with a behavior for the  decay of the central height of the peak(s) of $\CP$, which is found to be  proportional to $t^{-0.55}$. This can be detected clearly from Fig. \ref{corr-noncon-momentum} (b). In order to  double-check this non-diffusive behavior of the momentum self-diffusion, we plot the rescaled side peaks of $\CP\cdot t^{\gamma}$ in a co-moving frame at the sound velocity $c$ for different times: in Fig. \ref{corr-noncon-momentum} (c) with $\gamma=0.50$ (diffusive) and in (d) with $\gamma=0.55$ (non-diffusive). It is fair to say that the value $\gamma=0.55$ fits much better the data. This in turn indicates that the self-diffusion behavior of the momentum is  non-diffusive for the symmetrically amended rotator model at zero pressure. For this model, the momentum autocorrelation function $\CP$ coincides with the two sound normal modes $f_{\pm1}$ defined in NFHT. However, our numerical results of $\gamma=0.55$ again deviates from the prediction that $\gamma=0.50$ from NFHT \cite{Spohn2014JSP}.

\subsection{Hard point gas model with a square well potential and alternating masses}
The hard point gas model mimics a sort of idealized fluid with unbounded interactions strength. The Hamiltonian of a one-dimensional hard point gas model can be expressed as \cite{Mendl2014pre}:

\begin{equation}\label{ham-hpg}
H_{HPG}=\sum^{N}_{i=1}\frac{1}{2m_i}p^2_i+\frac{1}{2}\sum^{N}_{i\neq j=1}V(q_i-q_j)\;,
\end{equation}
where the setup of masses $m_i=1$ for even $i$ and $m_i=3$ for odd $i$, see in Ref. \cite{Mendl2014pre}. This choice converts this model into a non-integrable dynamics with strong mixing properties. The latter aspect is advantageous when it comes to the convergence issues at long times and large sizes in MD simulations. The symmetric square-well interaction potential reads \cite{Mendl2014pre}
\begin{equation}
V_{sw}(x)=0,\, \mbox{if}\,\, 0<|x|<1;\,\,\,V_{sw}(x)=\infty,\,\mbox{otherwise}\;.
\end{equation}

Because each unit cell contains two particles, the local energy $H_j$ and the momentum $p_j$ used for calculation need to be redefined as $H_j=H_{2j-1}+H_{2j}$ and the local momentum as $p_j=p_{2j-1}+p_{2j}$ where the number of unit cells amounts to half of the total particles.

\begin{figure}
 \centering
\includegraphics[width=0.42\textwidth]{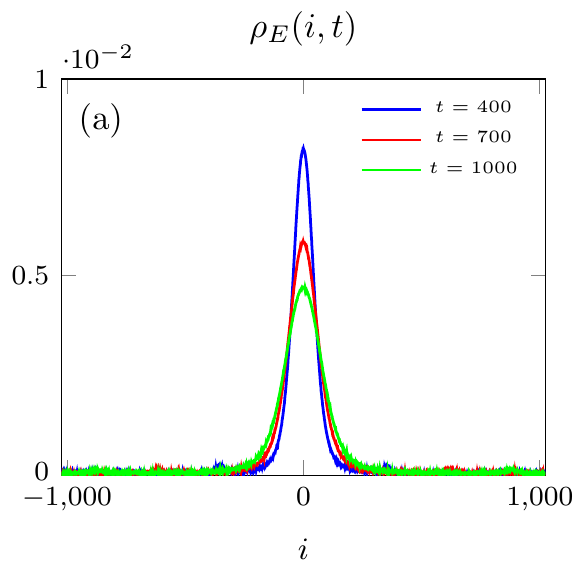}
\includegraphics[width=0.40\textwidth]{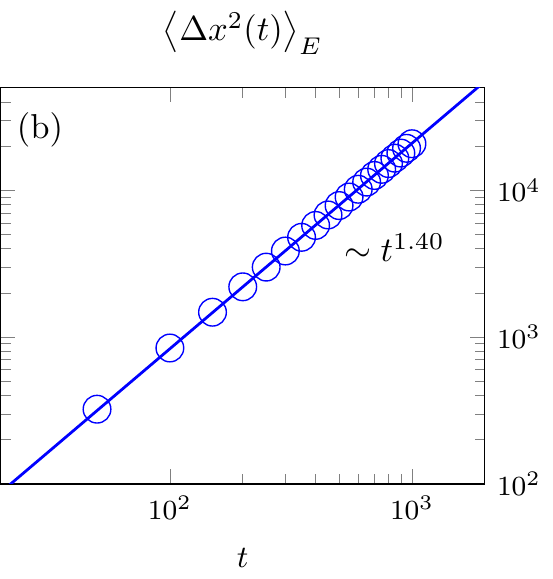}
\caption{(color online) Spreading of heat in the hard point gas model with symmetric square-well interaction potential with alternating masses at zero internal pressure: (a): The normalized energy correlation functions of excess energy density $\rho_E(i,t)=\CE$. The correlation times are $t=400$ (blue), $700$ (red), and $1000$ (green). (b): The anomalous MSD of the energy spread $\left<\Delta{x^2(t)}\right>_E$. The solid blue power law lines serve as a guide to the eye for the power law like behavior of the data in the asymptotic large time regime.  The parameters used in the numerical simulations are identical to the choice made  in Ref. \cite{Mendl2014pre} with a total number of particles  $N=4096$.}
\label{corr-hpg-energy}
\end{figure}

According to NFHT \cite{Spohn2014JSP}, this hard point gas model with a square well interaction potential and alternating masses can be classified into the same class as the FPU-$\beta$ lattice, and alike the amended coupled rotator model. In this model, the energy and momentum autocorrelation functions $\CE$ and $\CP$ coincide with the heat mode $f_0$ and sound modes $f_{\pm1}$ in the NFHT, respectively. The NFHT predicts that the energy diffusion is Levy walk  superdiffusive with $\left<\Delta x^2(t)\right>_E\propto t^{1.50}$, whereas the self-diffusion of momentum is predicted within NFHT to be normal diffusive.

In Fig. \ref{corr-hpg-energy} (a), we depict the energy autocorrelation function $\CE$ at different times. Compared with the amended rotator model and the FPU-$\beta$ lattice, the two side peaks are much smaller, although still not vanishing (being only barely visible in Fig. 4 (a)). The MSD of the energy spread is plotted in Fig. \ref{corr-hpg-energy} (b), yielding  a superdiffusive behavior with $\left<\Delta x^2(t)\right>_E\propto t^{1.40}$. As for the FPU-$\beta$ lattice and our amended coupled rotator model result our finding distinctly deviates from the NFHT prediction; it is however consistent with our numerical results of amended rotator model as well as the previously studied FPU-$\beta$ lattice, which all yield numerically an exponent $\beta= 1,40$. This again may indicate that NFH-Theory is quite good, although not sufficiently  accurate enough to account for the full nonlinear dynamics at work.

\begin{figure}
\centering
\includegraphics[width=0.40\textwidth]{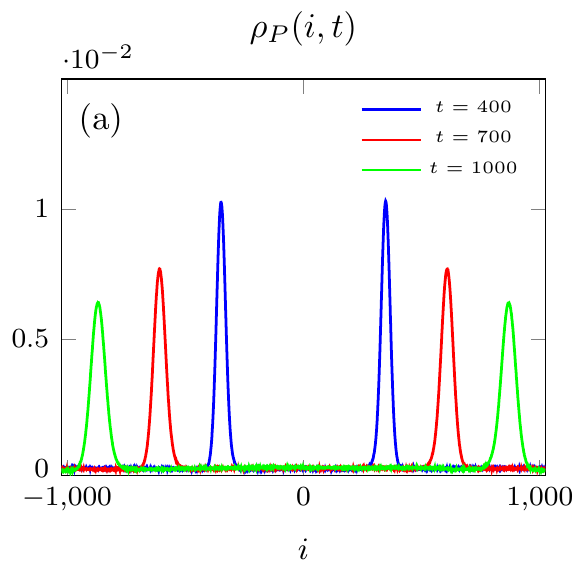}
\includegraphics[width=0.406\textwidth]{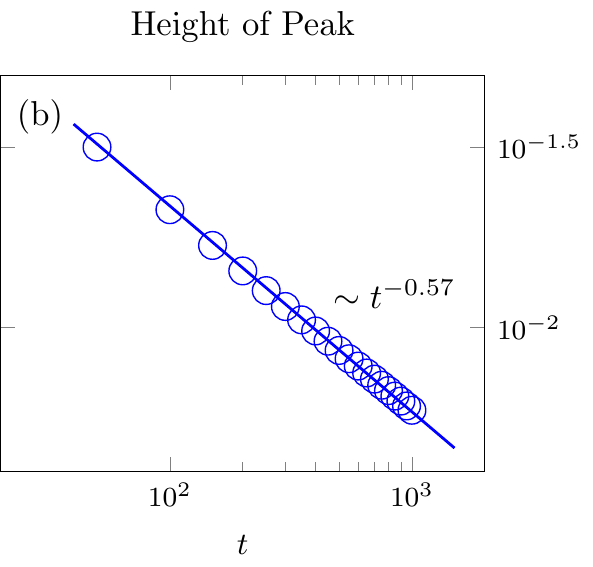}
\includegraphics[width=0.39\textwidth]{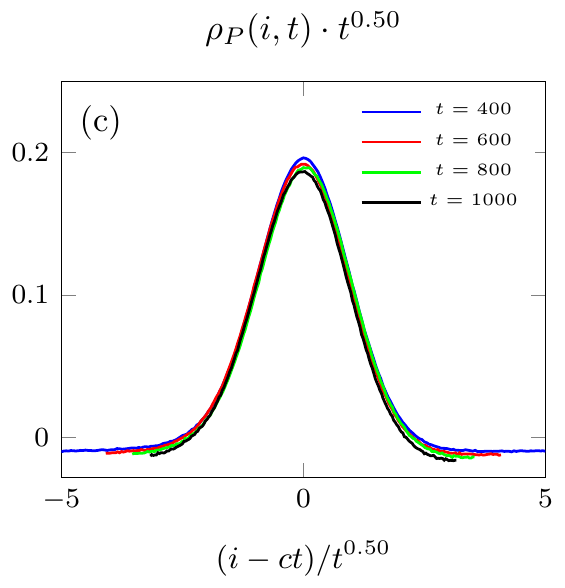}
\includegraphics[width=0.38\textwidth]{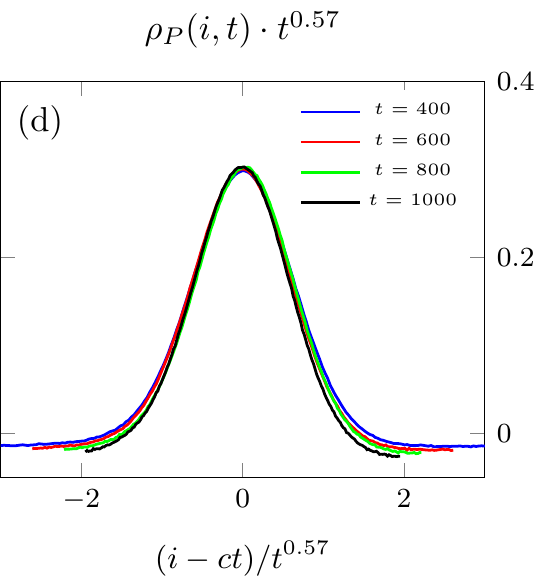}
\caption{Momentum spread in the hard point gas model with a square-well interaction potential composed of alternating masses at vanishing internal pressure: (a): The normalized correlation functions of excess momentum density $\rho_P(i,t)=\CP$. The correlation times are $t=400$ (blue), $700$ (red), and $1000$ (green). Each has two symmetric side peaks moving in opposite direction with a constant sound velocity $c$. (b): The decay of the height of the side peaks of $\rho_P(i,t)$. The solid blue power law lines depict a decay law proportional to $\sim t^{-0.57}$ as the best fit for the data from $t=400$ to $t=1000$. (c) The rescaled plot of the side peaks of $\rho_P(i,t)$ with the exponent  $0.50$ in the co-moving frame of the sound speed $c$ at $t=400, 600, 800$ and $1000$. (d)  The rescaled plot of the side peaks of $\rho_P(i,t)$ with the exponent of $0.57$ in the moving frame of sound velocity $c$ at $t=400, 600, 800$ and $1000$. The parameters used in the numerical simulations are the same as in Ref. \cite{Mendl2014pre} with $N=4096$.}
\label{corr-hpg-momentum}
\end{figure}

Of greater concern are the deviations for momentum spread which theory predicts to be normal but which seemingly does not fit  our numerical results.
The momentum autocorrelation functions $\CP$ at different times are depicted in Fig. \ref{corr-hpg-momentum} (a). Here we find  results that are quite similar to the amended coupled rotator model:  The two side sound peaks move in opposite direction with a constant sound speed $c$. To explore the momentum self-diffusion behavior in greater detail, we closely investigate the decay of the central height of the two side peaks, see in Fig. \ref{corr-hpg-momentum} (b). This  decay of the height of the peak are best fitted with a decay law proportional to $t^{-0.57}$. Being different from the normal diffusive scaling $t^{-0.5}$ this indicates  a non-diffusive behavior for the  momentum spread. The rescaled momentum excess density  $\CP\cdot t^{\gamma}$ in the co-moving frame of the sound velocity of the center of the side peaks are plotted in Fig. \ref{corr-hpg-momentum} (c) with  (i) $\gamma=0.50$ (normal diffusion) and also (d) with (ii) $\gamma=0.57$ (anomalous superdiffusion). Most importantly, the  curves with $\gamma=0.57$ fit convincingly better with the numerical data. This feature therefore reconfirms (contrary  to the NFHT prediction \cite{Spohn2014JSP,Mendl2014pre})  anomalous momentum spread for the hard point gas with a square well interaction potential.

\subsection{Testing a Lennard-Jones pair interaction}
Inspecting the preceding three test model cases one is led to speculate that it may well be the
unbounded part of the interaction potential that is at the cause
for a normal heat and momentum  transport behavior in nonlinear 1D  momentum-conserving
lattices. Such a reasoning  has obtained support in view of the recent numerical
studies by Savin and Kosevich \cite{Savin2014pre} which numerically find that
heat conductivity  remains finite in 1D interaction potentials possessing a
regime that allows  for dissociation at asymptotic large interaction distances
as it occurs, for example,  with the Lennard-Jones 1D case. If so, then for our hypothesis
to hold up we should find that in this case the subleading momentum self-diffusion behavior should emerge
 normal.

\begin{figure}
  \centering
\includegraphics[width=0.385\textwidth]{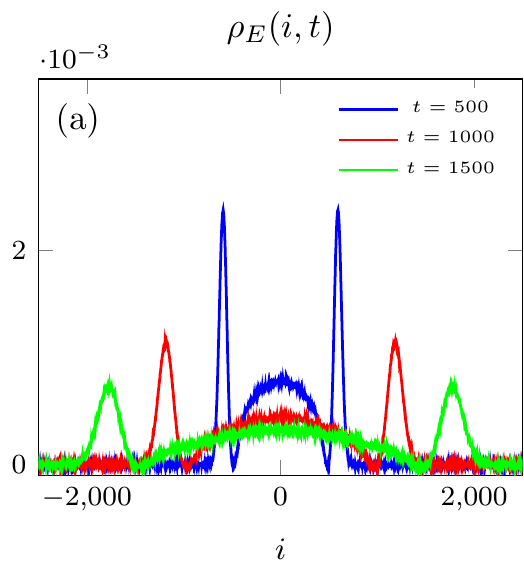}
\includegraphics[width=0.409\textwidth]{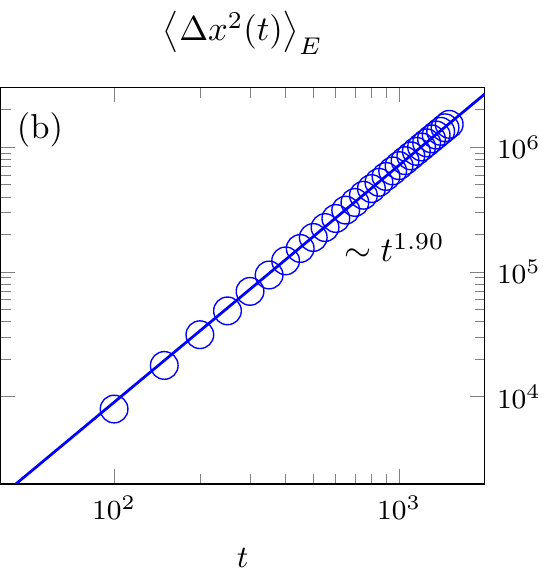}
\includegraphics[width=0.405\textwidth]{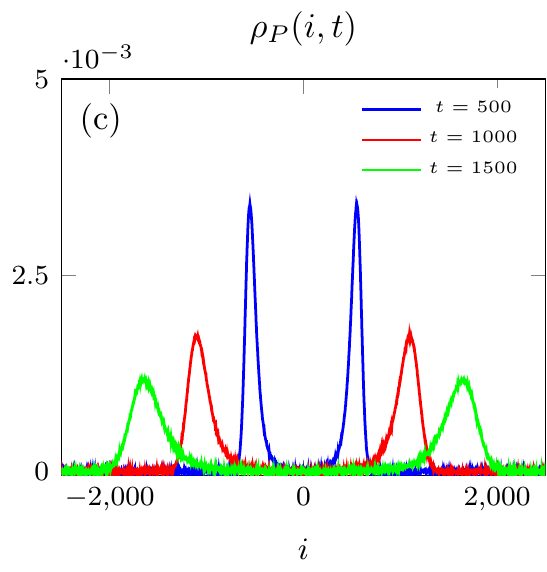}
\includegraphics[width=0.426\textwidth]{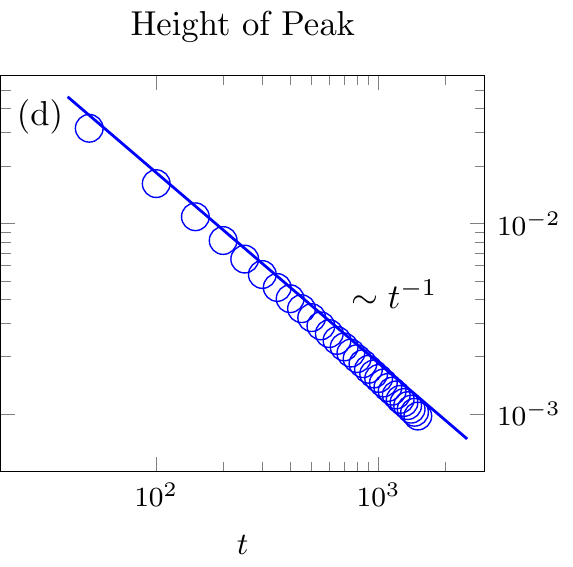}
  \caption{(color online) Energy spread in a 1D Lennard-Jones lattice system. (a) and (c): The normalized correlation functions of the excess energy density and
excess momentum density $\rho_E(i,t)=\CE$, $\rho_P(i,t)=\CP$  for the case with a Lennard-Jones interaction potential. The
correlation times are $t=500$ (blue),$1000$ (red), and $1500$ (green).
(b): The MSD of the energy spread $\left<\Delta{x^2(t)}\right>_E$. (d): The decay of the height of the side peaks of $\rho_P(i,t)$. In both situations (b) and (d), the solid blue power law lines serve as a guide to the eye for the data in the large time regime.  The parameters in the
numerical simulations are for $N=5001$,  $\sigma=2^{-1/6}$ and $\varepsilon=1/72$,
which are the same parameters as used in Savin and Kosevich's paper \cite{Savin2014pre}. The
calculated equilibrium temperature is at $T\approx 0.002$.}
\label{corr-nonconlj}
\end{figure}

Using the same numerical schemes as for the foregoing three lattice cases we next test our
hypothesis for a Lennard-Jones setup.
The corresponding Hamiltonian is given by
\begin{equation}\label{ham-lj}
  H =  \sum_{i} \left[ \frac{ p_{i}^2}{2} + 4\varepsilon \left(
(\frac{\sigma}{1+q_{i+1}-q_{i}})^6-\frac{1}{2}\right)^2 \right]  ,\
\end{equation}
using the same parameters as in Savin and Kosevich's paper; i.e., $\sigma=2^{-1/6}$ and
a binding energy $\varepsilon=1/72$ \cite{Savin2014pre}.
Here, the pair interaction potential is unbounded at short interaction distances
but becomes free at large  interaction distances, allowing dissociation. Due to this asymmetry in the interaction potential the internal pressure $\Lambda$ assumes  a finite value. The autocorrelation functions $\CE$ and $\CP$ for energy and
momentum are defined as before with Eqs. (\ref{e-corr}) and (\ref{p-corr}), respectively.

In Fig. \ref{corr-nonconlj} (a), we depict the  correlation functions $\CE$ for
the  energy diffusion
versus the correlation time $t$.  For sufficient large times
$t$ we observe that the
energy autocorrelation function $\CE$ evolves with two broadened side peaks,
being  rather distinct from a normal, Gaussian-like energy distribution spreading.
Consequently, the corresponding energy MSD is therefore  not normal, i.e. it is not proportional to time $t$. In
fact it assumes at  long times a power-law like behavior, being
below an overall  ballistic spreading, cf. Fig.  \ref{corr-nonconlj} (b).

Let us next also study the momentum spread  for this test case. In Fig. \ref{corr-nonconlj} (c), the momentum autocorrelation function $\CP$ at different times are shown. The decay of the height of the side peaks are also depicted with   Fig. \ref{corr-nonconlj} (d). We detect numerically a  behavior for the decay of the peak heights proportional to  $t^{-1}$. In perfect agreement with our stated  hypothesis, we thus find  as well  a non-diffusive momentum self-diffusion for this forth test case. Our findings not only contradict the recent results reported with \cite{Savin2014pre}, predicting therein a normal behavior for heat transport,
but as well make evident that it is {\it not} necessarily  the shape of the interaction potential which rules whether transport proceeds normal or anomalous.

\section{Conclusions and outlook}

The objective of studying energy and momentum transport in low-dimensional systems has recently attracted renewed interest  in view of  profound advances in theory, namely (i) the derivation of new transport relations \cite{Liu2014prl} and (ii) new  insight into scaling behaviors \cite{Das2014pre,Mendl2013prl,Spohn2014JSP,Mendl2014pre,Liu2014prb}. Apart from the role of energy spread and energy transport also the  problem of associated momentum spread and momentum transport gained recent attention \cite{Spohn2014JSP,Zhao2006prl,yunyun,Spohn2014arxiv}. Despite this recent progress many open problems  remain and the regime of validity of approximate theory predictions, most prominently for the appealing nonlinear fluctuating hydrodynamics  theory \cite{Spohn2014JSP}, is still under active debate.

With this work we  studied transport and  diffusion characteristics of different classes of momentum-conserving nonlinear $1D$ Hamiltonian dynamics for  both, heat and momentum. Using recent results of Ref. \cite{Liu2014prl} we started out showing that for energy
diffusion there exists a close relationship between the behavior of excess
energy diffusion  and the overall conductivity behavior for thermal heat
transport. This relationship has then been generalized alike for the case of
momentum diffusion in 1D nonlinear lattices.  For the subleading part of momentum spread beyond its possible ballistic transport  yields a diffusivity which relates to the time derivative of the asymptotic MSD for excess momentum, see in Eq. (\ref{eq:D_P}).
The consideration of momentum spread offers the possibility to quantify an
effective viscosity, being proportional to the momentum diffusivity, Eq. (\ref{eq:eta-D_P}). For normal
momentum diffusion  this effective viscosity is finite while it diverges with increasing time $t$ if
the intrinsic momentum diffusion occurs superdiffusive.

A main open problem in this field is the question when and under what conditions the energy and momentum transport deviate from normal. Put differently, when is transport and diffusive spreading occurring anomalously in low dimensional nonlinear Hamiltonian systems. -- In this context  the authors here put forward their speculative hypothesis that normal (anomalous) heat transport has its origin in normal (anomalous) momentum spread, and vice versa. Having no proof available for this hypothesis we tested the claim by investigating numerically  four different nonlinear model systems of momentum conserving nonlinear dynamics that are expected to belong to different classes for their energy/momentum transport characteristics. These were (i) the coupled rotator dynamics, (ii) its generalization involving the addition of unbounded harmonic interactions, (iii) the hard point gas and (iv)  a case with an asymptotic free dissociation regime (Lennard-Jones interaction potential).

As a main finding from these extensive numerical simulations we can assess that our so stated hypothesis does hold up. This encouraging positive result, however, does not assure that it is fundamentally correct,  as we have tested only a finite sample of nonlinear Hamiltonian models. Moreover, one may argue fairly that any numerical verification lacks a profound analytical foundation. Particularly, the question remains whether the numerical findings still hold true in  the extreme asymptotic regime of time $t\rightarrow \infty$, being beyond any numerical accessibility at this time. It can be convincingly stated, however, that the mere conservation of momentum in 1D Hamiltonian systems does generally not imply anomalous transport.

Our simulations also shed new light on the question of whether the recent NFHT \cite{Spohn2014JSP} is accurate enough to predict  the scaling regimes for energy and momentum transport. As mentioned, this theory is approximative in that it is based on a expansion of the Euler equations to second order only. In addition, it involves further approximations such as a decoupling of different modes at large times, which seemingly cannot be convincingly justified in presence of anomalous, superdiffusive  energy transport. Nevertheless, this theory admittedly is the best available at present times. Its scaling prediction for energy transport in  models with symmetric unbounded interaction potentials yields an exponent $\beta= 0.50$; this being  quite close, but  still distinctly different from our numerical value that $\beta=0.40$. Even more interesting is the prediction of NFHT that momentum spread should occur normal in these cases, thus violating our stated hypothesis. Our precise numerics shows however that such a normal momentum diffusion behavior does not fit with our  numerical findings. This has  been shown with the non-diffusive decay characteristics of the central peaks of the two opposite moving two side peaks in the excess momentum density function. This deviation is additionally substantiated  with the failure of a collapse of the data for an assumed normal diffusion in the co-moving  frame of  sound propagation. The behavior rather fits beautifully, however, with a collapse  using  {\it anomalous} momentum diffusion; -- thereby corroborating our stated hypothesis. In this context we may point out  that similar deviations from a normal diffusive scaling for the sound mode are present  in the numerics performed by the advocates of NFHT: upon inspecting  Fig. 8 in Ref. \cite{Das2014pre} one detects a similar failure of a diffusive collapse. The numerically established failure here of a diffusive collapse  for the case of the fully chaotic hard point gas is particularly trustworthy as we profit from underlying fast numerical converge features.

An interesting question for future studies is whether the criterion
can be extended to anomalous/normal heat flow occurring in two-dimensional
momentum-conserving nonlinear lattice systems. Typically, the anomalous heat conductance  then tends to diverge  in system size logarithmically \cite{Lepri2003pr,Dhar2008ap,xu2014nc,Lippi2000jsp,Yang2006pre,Xiong2010pre,Wang2012pre}.
Last but not least, the discussed complexity of normal versus anomalous heat and momentum transport in low
dimensions might possibly be put to constructive use when designing 1D low
dimensional devices for function, such as it is the case for the timely topic of
``phononics'' \cite{LIrmp}.

\section{Acknowledgements}
The authors much appreciate those many mutually stimulating and insightful correspondences with Professor Herbert Spohn. We also appreciate the useful correspondence with Professor Abhishek Dhar and his collaborators. The numerical calculations were carried out at Shanghai Supercomputer Center, which has been supported by the NSF China with grant No. 11334007 (B.L.). This work has been supported by the NSF China with grant No. 11334007 (Y.L., N.L., B.L.), the NSF China with Grant No. 11205114 (N.L.), the Program for New Century Excellent Talents of the Ministry of Education of China with Grant No. NCET-12-0409 (N.L.), the Shanghai Rising-Star Program with grant No. 13QA1403600 (N.L.), the NSF China with grant No. 11347216 (Y.L.) and Tongji University grant No. 2013kJ025 (Y.L.).

\section{Appendix}
\subsection{Dimensionless units}
For the investigation of the dynamics of 1D nonlinear lattice models, dimensionless units  have been applied throughout  as a convenient tool. As discussed in Ref. \cite{LIrmp}, the setup of dimensionless units is model dependent. We will elaborate below the details of the used dimensionless units for the 1D nonlinear lattice models considered in this work.

\subsubsection{Coupled rotator model.}
The dimensional Hamiltonian of coupled rotator model can be expressed as
\begin{equation}\label{dim-ham-cr}
H=\sum_i\left(\frac{p^2_i}{2m}+V\left[1-\cos{\frac{2\pi(q_{i+1}-q_i)}{a}}\right]\right)\;,
\end{equation}
where  $p_i$ and $q_i$ denote the dimensional momentum and displacement from equilibrium position for $i$-th atom. $m$ denotes the atom mass and $a$ is the lattice constant. The parameter $V$, possing the dimension of energy, represents the coupling strength of the neighboring rotators.

For this coupled rotator model, one can introduce the dimensionless variables by measuring lengths in units of $[a/(2\pi)]$, energies in units of $[V]$, masses in units of $[m]$, momenta in units of $[(Vm)^{1/2}]$, time in units of $[am^{1/2}/(2\pi V^{1/2})]$. The temperature will be measured in units of $[V/k_B]$ where $k_B$ is the Boltzmann constant. If we implement the following substitutions:
\begin{equation}\label{trans-rotator}
H\rightarrow H[V],\,\,p_i\rightarrow p_i[(Vm)^{1/2}],\,\,q_i\rightarrow q_i[a/(2\pi)]\;.
\end{equation}
The Hamiltonian of Eq. (\ref{dim-ham-cr}) can be transformed into the dimensionless one of Eq. (\ref{Ham-con}).

\subsubsection{Amended coupled rotator model.}
The dimensional Hamiltonian of amended rotator model is
\begin{equation}\label{dim-ham-ar}
H=\sum_i\left(\frac{p^2_i}{2m}+V\left[1-\cos{\frac{2\pi(q_{i+1}-q_i)}{a}}\right]+\frac{k_0}{2}(q_{i+1}-q_i)^2\right)\;,
\end{equation}
where $k_0$ denotes the extra coupling strength between neighboring atoms. The dimensionless units setup is the same as that for coupled rotator model. Applying the same transformation of Eq. (\ref{trans-rotator}),  Eq. (\ref{dim-ham-ar}) can be transformed into the dimensionless Hamiltonian of Eq. (\ref{ham-ar}) with the dimensionless $K=a^2k_0/(4\pi^2V)$.

\subsubsection{Hard point gas with alternating masses with square well potential.}
The dimensional Hamiltonian of hard point gas model is
\begin{equation}\label{dim-ham-hpg}
H=\sum^N_i\frac{p^2_i}{2m_i}+\frac{1}{2}\sum^N_{i\neq j=1}V(q_i-q_j)\;,
\end{equation}
where $m_i$ is the mass for $i$-th particle and the square well potential can be described as
\begin{equation}
V_{sw}(x)=0,\, \mbox{if}\,\, 0<|x|<a;\,\,\,V_{sw}(x)=\infty,\,\mbox{otherwise}\;,
\end{equation}
with $a$ denoting the average distance between neighboring particles. The alternating masses are introduced by setting particle masses $m_i=m_0$ for an even number of $i$ and $m_i=3m_0$ for an odd number of $i$.

For this hard point gas model, one can introduce the dimensionless variables by measuring lengths in units of $[a]$, masses in units of $[m_0]$. Since there is no characteristic potential energy for this model, its dynamics is essentially the same for any energy scale. One can arbitrarily choose an energy scale $E_0$ as the reference energy and the energies can be measured in units of $[E_0]$. As a result, the momenta can be measured in units of $[(m_0 E_0)^{1/2}]$ and the time can be measured in units of $[a(m_0/E_0)^{1/2}]$. The temperature can also be measured in units of $[E_0/k_B]$. In our study we used the same parameters as used in Ref. \cite{Mendl2014pre}.

\subsubsection{Lennard-Jones model.}
The dimensional Lennard-Jones model has the following Hamiltonian
\begin{equation}\label{dim-ham-lj}
H=\sum_i\left[\frac{p^2_i}{2m}+4\varepsilon\varepsilon_0\left(\left(\frac{\sigma}{1+(q_{i+1}-q_i)/a}\right)^6-\frac{1}{2}\right)^2\right]\;,
\end{equation}
where $m$ is the atom mass and $a$ is the lattice constant. $\varepsilon\varepsilon_0$ denotes the binding energy and $\varepsilon$ is a dimensionless parameter. $\sigma$ is yet another dimensionless parameter.

For this Lennard-Jones model, one can introduce the dimensionless variables by measuring lengths in units of $[a]$, masses in units of $[m]$, energies in units of $[\varepsilon_0]$, momenta in units of $[(\varepsilon_0 m)^{1/2}]$, time in units of $[a(m/\varepsilon_0)^{1/2}]$. The temperature will be measured in units of $[\varepsilon_0/k_B]$. If we implement the following substitutions
\begin{equation}\label{trans-lj}
H\rightarrow H[\varepsilon_0],\,\,p_i\rightarrow p_i[(\varepsilon_0 m)^{1/2}],\,\,q_i\rightarrow q_i[a]\;.
\end{equation}
The Hamiltonian of Eq. (\ref{dim-ham-lj}) can then be transformed into the dimensionless Hamiltonian of Eq. (\ref{ham-lj}).

\subsection{Numerical procedures}
In order to obtain  precise numerical results, we  employ
MD simulations for an isolated system evolving with the corresponding Liouvillian  over large,
extended time spans and used throughout periodic boundary conditions. The method to obtain the
correlation functions is adopted from Ref. \cite{Chen2013pre}.
The equations of motions are integrated with a fourth order symplectic algorithm
\cite{Laskar2001cmda,Skokos2009}.

\end{document}